\documentclass[a4paper,12pt]{article}

\usepackage{graphicx}

\newcommand{\beq}{\begin{eqnarray}}
\newcommand{\eeq}{\end{eqnarray}}

\newcommand{\kbar}{$\bar{K}$}

\begin{document}

\begin{center}
{\large \bf
Emission spectra and invariant masses of $\Lambda$ and $p$ in the stopped-$K^-NN$ absorption process in $^4$He and $^6$Li
}

\vspace{1cm}
(NPA  resubmission, 2007-5-12 )

\vspace*{0.5cm}
Toshimitsu Yamazaki$^{a)}$ and Yoshinori Akaishi$^{b)}$\\
 \vspace*{2mm}
 {\it a) Department of Physics, University of Tokyo, Bunkyo-ku, Tokyo 113-0033, Japan, and
RIKEN, Wako, Saitama 351-0198, Japan, b) College of Science and Technology, Nihon University, Funabashi, Chiba 274-8501, Japan, and RIKEN, Wako, Saitama 351-0198, Japan}
\end{center}

\begin{abstract}
We have calculated the emission spectra of $Y$ and $N$ and invariant masses of $YN$ pairs in the direct $K^-NN \rightarrow YN$ absorption process at rest in $^4$He and light nuclei in order to provide theoretical tools for correct interpretations of experimental data with or without invoking kaonic nuclear bound states. All the momentum distributions are broad with widths around 150-200 MeV/$c$ (except for the case of $^6$Li target), while the partial invariant mass of each $YN$ pair has a peak around 2310-2330 MeV/$c^2$. We argue against the interpretations of stopped-$K^-$ experimental data of KEK and FINUDA by Oset and Toki and by Magas, Oset, Ramos and Toki.
\end{abstract}

\begin{quotation}
\end{quotation}


\section{Introduction}

In recent years we have predicted deeply bound kaonic nuclear states in light nuclei (often called $\bar{K}$ nuclear clusters, {\it KNC}) and studied their structure and formation \cite{Akaishi:02,Yamazaki:02,Dote:04a,Dote:04b,Yamazaki:04,Akaishi:05,Kienle:06}, using bare \kbar $N$ interactions deduced based on the empirical data of \kbar $N$ systems \cite{Akaishi:02}. 
The most characteristic feature of these \kbar~ bound states is that the strongly attractive \kbar $N$ interaction of $I=0$ channel causes shrinkage of nuclei and often helps to form bound states on non-existing nuclei, such as $K^-pp$ \cite{Yamazaki:02}. There are some experimental reports \cite{Iwasaki:03,Suzuki:04,Iwasaki:06,FINUDA-PRL,Kishimoto:05,Aslanyan}, which are related to the issue on the existence of such bound states, but no conclusive evidence has been established so far. Whereas more dedicated experiments are necessary in the future, as being planned at KEK, DAPHNE, GSI and others, fair and useful theoretical framework to interpret existing (and also future) experimental data on unbiased bases is of vital importance. The present paper is thus aimed at providing theoretical emission spectra of  hyperons and nucleons and their relevant invariant masses in the $K^-$ absorption process, both with and without invoking bound state formation. By comparing any experimental data with those  ``background processes" one can judge without prejudice whether an ``exotic state" is indicated or not.  We have developed valid theoretical tools to interpret experimental data of $K^-$ absorption at rest, and thereby to help experimentalists to deduce fair and correct implications of their data. It is needless to say that an incorrect assessment of a background process would cause a wrong conclusion concerning the signal that one is looking for.

A recently published paper by Oset and Toki (hereafter called OT) \cite{Oset:05} not only criticizes the basic ingredients of Akaishi and Yamazaki \cite{Akaishi:02} (to which we have given counter arguments elsewhere \cite {Akaishi:06}), but also made their own interpretation on the existing experimental data. From their "recoilless" absorption treatment of $[K^- pN]_{\rm atom} \rightarrow Y + p$, they predicted discrete lines at 488 and 562 MeV/$c$ in the proton spectrum of $^4$He(stopped-$K^-, p$), and thereby ascribed the 500 MeV/$c$ peak that was reported by KEK E471 \cite{Suzuki:04} to this origin. In the mean time, a new experiment E549 at KEK \cite{Iwasaki:06} has revealed no such discrete peaks in an inclusive proton spectrum. In another paper Magas, Oset, Ramos and Toki (hereafter called MORT) \cite{MORT} calculated the invariant-mass spectrum of  $\Lambda$ and $p$ in the  stopped-$K^-$ process in $^7$Li, and insisted that the final-state interaction (FSI) of $\Lambda$ and/or $p$ produces a large  bump in the $\Lambda$-$p$ invariant-mass spectrum, leaving the initially formed $M = 2340$ MeV/$c^2$ component very little. They claimed that the FINUDA spectrum \cite{FINUDA-PRL} should be interpreted by such a mechanism without invoking a nuclear $K^-pp$ bound state. In this paper we argue against the procedures of OT and MORT. 

\section{Three different mechanisms in $\bar{K}$ absorption at rest}

We discuss the following three different cases, as shown in Fig.~\ref{fig:Kabs-mechanism}. This is for the case of $^7$Li, but can be applied to $^4$He as well. We compare the predicted quantities (momentum spectra and invariant masses of emitted $\Lambda$ and nucleons) with experiments of old bubble chambers \cite{Katz:70} and recent KEK \cite{Suzuki:04} and FINUDA \cite{FINUDA-PRL}. Our main purpose is to provide experimentalists with theoretical results to help their analyses of  future exclusive experiments including AMADEUS \cite{AMADEUS}.  \\

\noindent 
{\bf (A) OT: Recoilless $K^-NN$ absorption} \\

OT claim that {\it this kaon absorption process should happen from some $K^-$ atomic orbits which overlap with the tail of the nuclear density and hence the Fermi motion of the nucleons is small}. This is equivalent to postulating the following decay process: 
\begin{eqnarray}
[K^- \,^4{\rm He}]_{\rm atom} &=& [K^-NN]_{\rm atom} + [NN]_{\rm dead}, \label{eq:KNN-OT-1} \\
&&[K^- NN]_{\rm atom} \rightarrow Y + p,\label{eq:KNN-OT-2}
\end{eqnarray}
\noindent
where $[K^-NN]_{\rm atom}$ expresses a grouping of the atomic $K^-$ with two nucleons in $^4$He without coupling to another  $NN$ pair {\it at rest}, which we call $[NN]_{\rm dead}$ symbolically. They claim the mass of this subsystem  to be
\begin{equation}
M([K^-NN]_{\rm atom}) = M_0 - B_N/c^2 = 2342~{\rm MeV}/c^2,\label{eq:OT-MKNN}
\end{equation}
\noindent
which is less by a nuclear binding energy, $B_N$, than the free constituent mass:
\begin{equation}
M_0 = m_K + 2 \,M_p = 2370~{\rm MeV}/c^2.
\end{equation}
In the OT model the $[K^-NN]_{\rm atom}$ subsystem is an on-mass-shell object, and they attribute the $\alpha$ binding energy to the $B_N$ (namely, $B_N = B_{\alpha} = 28.3$ MeV), although they claim that the Fermi motion of the nucleons at $K^-$ capture is small.

In this model the proton and the hyperon are emitted back to back with equal discrete momenta. 
Magas {\it et al.} (MORT) \cite{MORT} took into account FSI of these originally monoenergetic particles of  back-to-back correlation with nucleons inside the remaining nucleus (in the case of $^7$Li target, it is $^5$H). They postulate the FSI effect to be so large ($\sim 90 \%$) as to diminish the original invariant-mass peak at 2342 MeV/$c^2$,  (\ref{eq:OT-MKNN}), to $\sim 10 \%$,  and as a result $M_{\rm inv} (\Lambda p)$ is widely distributed and the angular correlation is also spread out. \\

\noindent
{\bf (B) Direct QF $K^-NN$ absorption}\\

In our model, the subsystem $[K^-NN]_{\rm atom}$, decaying to  $Y + N$, is regarded as an off-mass-shell object  which is coupled to  the rest of the nucleus, bound in $[K^- \,^4{\rm He}]_{\rm atom}$. The whole absorption process, hereafter called {\it Direct Quasi-Free}, is
\begin{eqnarray}
[K^- \,^4{\rm He}]_{\rm atom} &\approx& [K^-NN]_{\rm atom} \, [NN] \nonumber\\
                       &\rightarrow& Y + N + [NN]', \label{eq:QF-KNN-2}
\end{eqnarray}
where the remaining $[NN]'$ is a spectator with a certain momentum. {\it Without invoking FSI} the momentum spectra of emitted $\Lambda$ and $N$ are not monoenergetic, and their angular correlation is smeared out to some extent. The ``invariant mass" distribution to be reconstructed from observed $\Lambda$ and $N$ momenta is close the mass limit, (\ref{eq:OT-MKNN}), but somewhat broad. This is categorized as a {\it Quasi Invariant Mass (QIM)}, which reflects the momentum distribution of a bound object, as discussed in detail for in-medium hadrons by Yamazaki and Akaishi \cite{Yamazaki:99}. No FSI effect is taken into account at the present stage of our calculation. \\

\noindent
{\bf (C) KNC: \kbar~nuclear cluster formation}\\

Next, we examine the following \kbar~ cluster formation process.
\begin{eqnarray}
[K^- \,^4{\rm He}]_{\rm atom} &\rightarrow& [K^-pp]_{\rm nucl} + n + n, \label{eq:KNC-AY-1} \\
&&[K^- pp]_{\rm nucl} \rightarrow \Lambda + p,\label{eq:KNC-AY-2}
\end{eqnarray}
\noindent
where $[K^-pp]_{\rm nucl}$ is a nuclear bound state with a binding energy $B_K$.  The invariant mass of $\Lambda-p$ is equal to the bound-state mass:
\begin{equation}
M_{\rm inv}(\Lambda p) = M( [K^-pp]_{\rm nucl}) = m_K + 2 M_p - B_K/c^2.
\end{equation}
\section{Calculation of $N$ and $d$ momenta in $^4$He}

Before going to the calculation of $\Lambda$ and $N$ emission spectra we calculated the momentum distributions of a $d$-like cluster ({\it quasi-d}) in $^4$He as well as in other light nuclei by using a $d$-core cluster model with the orthogonality condition. The momentum distributions of $(NN)$ cluster were investigated also microscopically by a variational method named ATMS \cite{ATMS}. It was shown that the momentum distributions are reasonably reproduced by the cluster model, except for high-momentum ($P_{(NN)} \geq 0.8~ {\rm GeV}/c$) components due to short-range repulsion and tensor force of the realistic $NN$ interaction, which, however, give only minor effects on the present problems.

The cluster-model results, which we have calculated in connection with the present paper, are shown in Fig.~\ref{fig:d-momentum}. The rms momentum of quasi-$d$ in $^4$He is 184 MeV/$c$, and it can never be close to zero. The calculated momentum distribution of $N$ in $^4$He is also similar, the rms momentum being 151 MeV/$c$. Later, we will see that this rms momentum is consistent with an experimental observation of the $\Lambda$ momentum in the $\Lambda + d + n$ final state in a bubble chamber experiment \cite{Katz:70} and a recent FINUDA experiment \cite{FINUDA-Fujioka}.

\section{Particle emission spectra in $K^-\,^4$He $\rightarrow p + Y + (NN)$} \label{sec:AY-pSigma}

We calculate the following quantities which can be observed experimentally: i) momentum spectrum of proton, ii)  momentum spectrum of $\Lambda$, and iii) angular correlation of $\Lambda$ and $p$, and iv) invariant-mass spectrum of $\Lambda$ and $p$. 


\subsection{Formalism}

We consider the proton and hyperon $Y^0$ momentum distribution in 
\begin{eqnarray}
&&[K^- \,^4{\rm He}]_{\rm atom} \rightarrow p + \Sigma^- + d,\label{eq:AY-KNN-1b}\\
&&[K^- \,^4{\rm He}]_{\rm atom} \rightarrow p + Y^0 + (nn).\label{eq:AY-KNN-1a}
\end{eqnarray}
In the limit of vanishing recoil (A) the proton momenta are given by discrete values:
\begin{eqnarray}
&&P_p = 483~{\rm MeV/}c~{\rm for}~Y=\Sigma^0, \label{eq:OT-value-Sigma0}\\
&&P_p = 561~{\rm MeV/}c~{\rm for}~Y=\Lambda. \label{eq:OT-value-Lambda}
\end{eqnarray}

First, we take up the Direct QF process (B), (\ref{eq:QF-KNN-2}), without invoking KNC formation (C). We derived the following formulae for the $K^-$ capture from the 2$p$ atomic orbit. In the case of $d$ emission (\ref{eq:AY-KNN-1b}) the decay spectrum is given by

\begin{eqnarray}
\frac{{\rm d}^2 \Gamma}{{\rm d} P_p \, {\rm d} P_Y} &=& C'\, P_p^2 \,P_Y^2 \, \mid F(P_d) \mid ^2, \nonumber \\
 F(P_d) &=& \int_0^\infty dr_K r_K^2\, \frac {U_{dd}(2r_K)}{2r_K}\, j_1(P_d 2r_K)\, r_K {\rm exp}(-r_K/a_{\rm B})
\end{eqnarray}
for kinematically allowed momenta of $p$ and $Y$ ($P_d$ is determined from momentum and energy conservations), where $U_{dd}$ is the radial wavefunction of $d-d$ relative motion in $^4$He. In our treatment the $K^-$ interacts  first with a single nucleon, and successively with a second nucleon, whereas in MORT the $K^-$ interacts with two nucleons at the same place.  

In the case of $(nn)$ emission (\ref{eq:AY-KNN-1a}) the $nn$ pair can carry some amount of its internal energy, which we estimate by using a harmonic oscillator model for $^4$He. Then, the following emission spectrum of $p$ and $Y$ is obtained:

\begin{equation}
\frac{{\rm d}^3 \Gamma}{{\rm d} P_p \, {\rm d} P_Y \, {\rm d} x} = C''\, P_p^2 \,P_Y^2 \, P_{(pY)}^2 \, {\rm exp}[-\frac{1}{a} (\frac{P_{(pY)}}{\hbar})^2] 
\, p_{nn} \, {\rm exp}[-\frac{2}{a}(\frac{p_{nn}}{\hbar})^2]. \label{eq:CS-1}
\end{equation}
The parameter $a$ for the nucleon binding given by the harmonic oscillator,
\begin{equation}
\hbar \omega = \frac{\hbar^2}{M_N} a = 21.6 \, {\rm MeV},
\end{equation}
represents the realistic wavefunction. The kinematical constraints among the various momenta are given by
\begin{eqnarray}
P_{(pY)}^2 &=& P_p^2 + P_Y^2 + 2 P_p P_Y \, x ~=~ P_{(nn)}^2, \label{eq:CS-2} \\
\frac{p_{nn}^2}{M_n} &=& M_p c^2 +m_K c^2 - M_Y c^2 - B_{\alpha}
-\frac{P_{(pY)}^2}{2 \mu_{(p Y)(nn)}}  \nonumber \\
&& -\frac{1}{2 \mu_{pY}} \frac{M_Y^2 \, P_p^2 + M_p^2 \,P_Y^2 - 2 M_p\, P_p \, M_Y\, P_Y\, x}{(M_p + M_Y)^2}, \label{eq:CS-3}
\end{eqnarray}
where $x={\rm cos\,}\theta_{pY}$. The subscripts $(pY)$ and $(nn)$ denote $p$+$Y$ and $n$+$n$ systems, respectively.

\subsection{Proton spectrum}\label{sec:proton-spectrum}

The $P_p$ spectrum in the case of $Y = \Sigma^0$ with a realistic $\alpha$ particle density ($\hbar \omega = 21.6$ MeV) is shown in Fig.~\ref{fig:P_p} (A). It has a broad distribution, as expected. On the other hand, the OT Ansatz claims the $(nn)$ or $d$ momentum to be close to zero and thus the momentum of $p$ to be discrete ($\sim 483$ and 561 MeV/$c$ for $Y = \Sigma^0$ and $\Lambda$, respectively). In the same figure we present the Dalitz domains which show kinematically allowed regions of the observable quantities. The realistically calculated proton distribution occupies a part of the domain, whereas the OT points are located in its extreme limits.

A very recent result of the proton spectrum in an improved stopped-$K^-$ experiment in $^4$He at KEK \cite{Iwasaki:06} has turned out to be in good agreement with this ``background process". The inclusive proton spectrum shows no discrete peak. 

\subsection{$\Lambda$ spectrum}

\begin{figure}
\centering
\includegraphics[width=12cm]{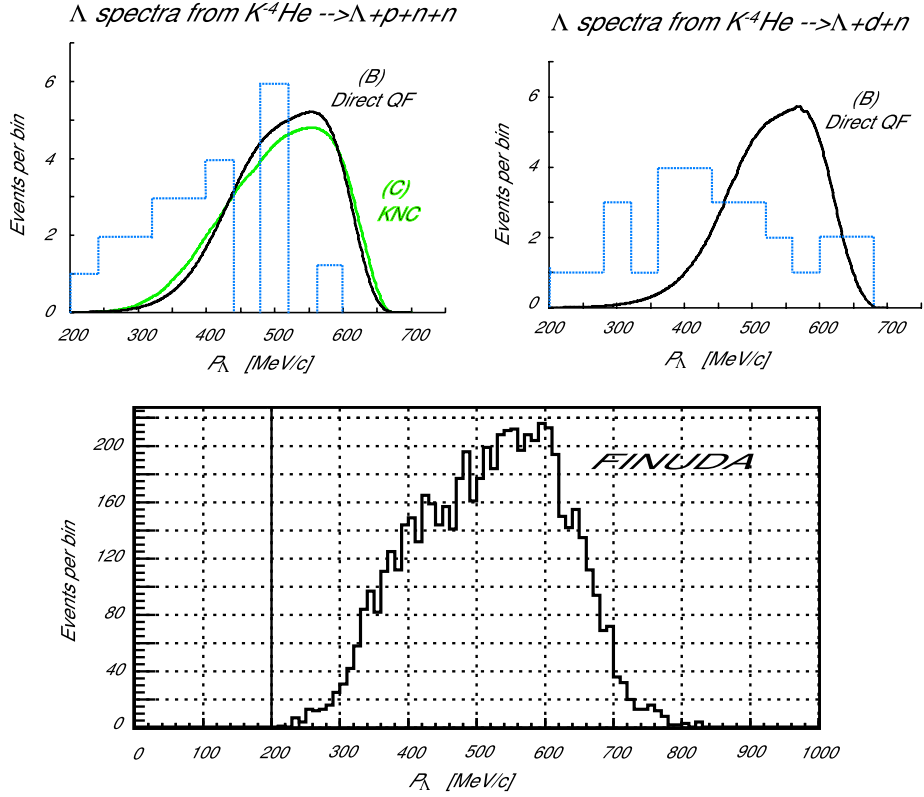}
\vspace{0cm}
\caption{\label{fig:P(Lambda)-Katz} 
(Upper) The calculated momentum distributions of $\Lambda$ in the cases of Direct QF (B) and KNC (C) $K^-pp \rightarrow \Lambda + p$ processes in $K^- + ^4$He. They are compared with old bubble chamber data for the exclusive decay channels of $\Lambda + n + d$ and $\Lambda + p + n + n$ \cite{Katz:70}. The Oset-Toki prediction corresponds to a discrete line at $P_\Lambda$ = 562 MeV/$c$.  (Lower) An inclusive $\Lambda$ momentum spectrum of FINUDA  from Li and C targets without acceptance correction, taken from \cite{FINUDA-Fujioka}, which agrees well with the theoretical distribution.}
\end{figure}

In Fig.~\ref{fig:P(Lambda)-Katz} we show the calculated momentum distribution of $\Lambda$ in the (B) Direct QF process. It distributes over a wide range of $400-650$ MeV/$c$. In the figure we compare the spectrum with an old bubble chamber data for $P_\Lambda$ with well selected final states ($\Lambda + d + n$ and $\Lambda + p + n + n$), which also shows a continuous spectrum. The higher momentum part is explained by our calculated curve, whereas the low-momentum part should be attributed to some other origins. 

This conclusion is further strengthened by a recent FINUDA data, which reveals with much higher statistics a very continuous $\Lambda$ spectrum without a trace of discrete lines \cite{FINUDA-Fujioka}. Although the FINUDA data were taken not from $^4$He but from light targets, the essential character of the two-nucleon absorption process should remain the same.

The $\Lambda$ momentum spectrum expected from the case of (C) KNC formation is also shown in the figure. Essentially, there is no distinction in $P_\Lambda$ between (B) and (C) cases. The case of KNC formation (C) is directly reflected in the $\Lambda$-$p$ invariant mass, as will be discussed in the next section.

\subsection{$K^-$ absorption density in $^4$He}

\begin{figure}
\centering
\includegraphics[width=10cm]{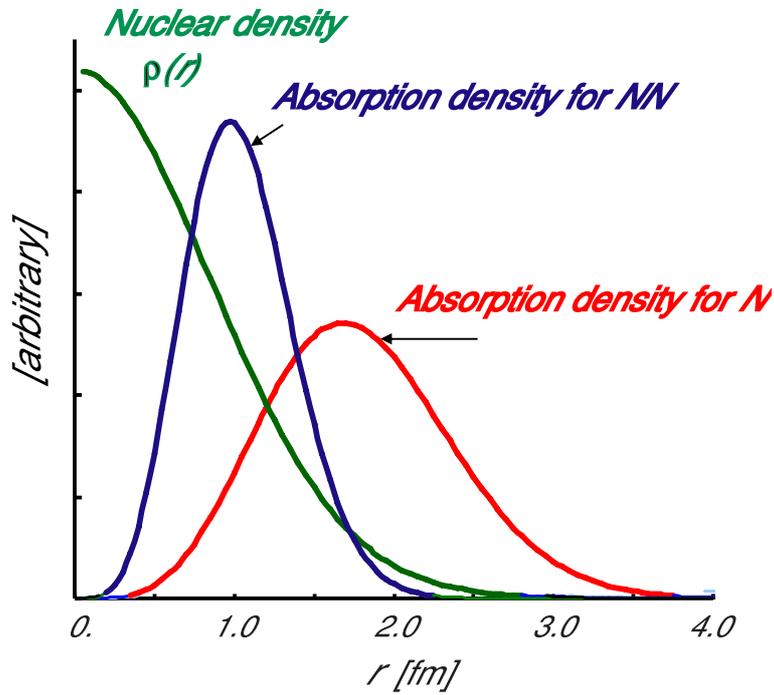}
\vspace{0.cm}
\caption{\label{fig:K-overlap} 
The $K^-$ absorption density distributions, when the $K^-$ in the $2p$ atomic state is absorbed by a single $N$ and by a pair of $NN$ in $^4$He.  The nuclear density distribution of $^4$He is also shown.}
\end{figure}

The momentum spectra we calculated are all broad with widths around 150-200 MeV/$c$. On the other hand, OT insist that, when a $K^-$ is captured at a remote end of the nuclear peripheral, the nucleon (and recoil) momenta become negligibly small so as to produce recoil-free discrete lines. To clarify this difference we calculated the ``absorption density" distribution of the absorbed $K^-$ and the nuclear density as a function of the $K^-$ coordinate $r_K$, defined as
\begin{equation}         
D(r_K) = {r_K}^2 |\Phi_{\alpha} (r_K)|^2 \, |\Psi_{2p} (r_K)|^2,\label{eq:absorption-density}
\end{equation}
using the realistic wavefunction for the 2$p$ orbital ($\Psi_{2p} (r_K)$) of $K^- \,^4$He atom in the presence and absence of the strong-interaction potential, and the $N$ distribution in $^4$He ($|\Phi_{\alpha} (r)|^2$). The result in the absence of the strong-interaction potential is shown in Fig.~\ref{fig:K-overlap} together with the nucleon density distribution $\rho (r) = |\Phi_{\alpha} (r)|^2$, which has an rms radius of 1.47 fm. The absorption density in the presence of the strong-interaction moves inward.  Although the $K^-$ wavefunction, given by $R_{2p} (r_K) = C\, r_K\, {\rm exp} (- r_K/a_{\rm B})$ in the outer region, spreads over the atomic scale with a mean radius of $<r_K>_{2p} = 155$ fm, the absorption density $D(r_K)$ has a distribution near the nuclear peripheral, centered around $r_K \sim 1.7$ fm. For comparison the two-nucleon absorption density by the treatment similar to MORT (namely, assuming the two $NN$ to be absorbed at the same place) is also shown. 

As we have shown above, the claim of OT for discrete momentum spectra is not physical. It violates the quantum mechanical law ($\Delta p \approx \hbar/\Delta x$), because the nuclear wavefunction of the ground state $\Phi_{\alpha}$ is responsible for the $K^-$ capture, no matter where the capture takes place. (One could, however,  conceive a situation where $K^-$ is captured by a peripheral nucleon, such as in the outermost orbital in a heavy nucleus; though this is not the case with $^4$He at all). 

\begin{figure}
\centering
\includegraphics[width=14cm]{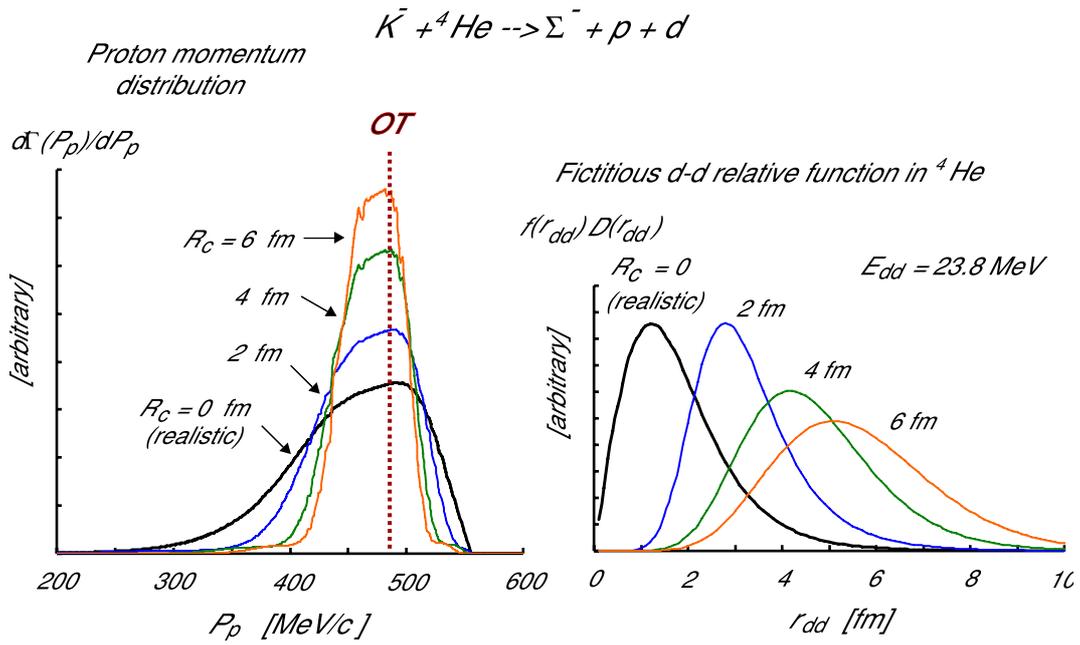}
\vspace{0cm}
\caption{\label{fig:Rc} 
(Left) The calculated momentum distributions of emitted proton in the cases of (B) direct QF $K^-pn \rightarrow \Sigma^- + p$ processes in $K^- + ^4$He, corresponding to the realistic case (black solid curve) and the cases of unrealistically filtered remote $K^-$ absorption, shown in (Right). The Oset-Toki prediction corresponds to a discrete line at $P_p$ =  482 MeV/$c$, designated by a broken vertical line.}
\end{figure}

From a pedagogic point of view, let us consider a fictitious case by artificially filtering the location $r_{dd}$ of $K^-$ absorption. For this purpose we imposed a smooth enough cutoff function, 
\begin{equation}
f(r_{dd}) =  [1 - {\rm exp}(-(\frac{r_{dd}}{R_C})^2)]^4,
\end{equation}
where $R_C$ is an artificial cutoff parameter for $r_{dd}$ caused by an arbitrarily (unrealistically) varied  $K^-$ absorption density distribution, as shown in the right-hand side of Fig.~\ref{fig:Rc}. The proton momentum spectra under such a treatment were calculated following the procedure to be described in the next section. As shown in the left-hand side of the figure, even in a remote capture ($R_C = 6$ fm) the proton spectrum remains essentially the same as in the realistic absorption case ($R_C = 0$). This comparison demonstrates that {\it the proton momentum distribution that reflects the nucleon momenta of $^4$He is nearly unchanged no matter how the location of the $K^-$ capture changes from inside to outside}. From all this examination we conclude that the $K^-NN$ absorption of the on-shell recoilless object  (\ref{eq:KNN-OT-2}), assumed by OT, cannot occur; substantial momenta should be transferred to the remaining nucleons. 


\section{Neutron spectra in $K^-\,^4$He $\rightarrow n + Y + (NN)$ and associated $Y$ decays} \label{sec:AY-nSigma}

\subsection{Hyperon-decay tagged neutron spectra}

In connection with the experimental trial of the KEK E471 and E549 experiments we discuss the neutron emission spectra in the $K^-$ capture in $^4$He at rest, associated with the direct capture processes as well as a possible tribaryonic bound state formation. The first kind is the neutron spectrum from the direct QF two-nucleon capture process, 
\begin{eqnarray}
K^-\, + ^4{\rm He} &\rightarrow& n + \Lambda (\Sigma^0) + (pn)~~[12 \%],\\
                                &\rightarrow& n + \Sigma^+  + (nn)~~[1\%],\\
                                &\rightarrow& n + \Sigma^- + (pp)~~[4 \%] ,
\end{eqnarray}
where the $(NN)$ denotes a spectator, and the numbers in the parenthesis are the estimated branching ratios. The branching ratios for the two-nucleon absorption processes are only partially known as 1.6 \% for $\Sigma^- pd$, 2.0 \% for $\Sigma^- ppn$ and 11.7 \% for $\Lambda (\Sigma^0) pnn$ without specifying the final state momenta from the old bubble chamber experiment of Katz {\it et al.} \cite{Katz:70}. 

The second kind is the neutron from the decay of hyperons which originate from the above direct process, 
\begin{eqnarray}
\Lambda &\rightarrow& p + \pi^-:~[64 \%],~~
                \rightarrow n + \pi^0:~[36 \%],\\
\Sigma^+ &\rightarrow& p + \pi^0:~[52 \%],~~
                  \rightarrow n + \pi^+:~[48 \%],\\
\Sigma^0 &\rightarrow& \Lambda + \gamma:~[100 \%],\\
\Sigma^-   &\rightarrow& n + \pi^-:~[100 \%].
\end{eqnarray}
We name the two kinds of neutron spectra, from the absorption process and from the decay process,  respectively, as
\begin{eqnarray}
&& F_n^{\Sigma-{\rm form}} (p_n), \\
&& F_n^{\Sigma-{\rm decay}} (p_n).
\end{eqnarray}

\begin{figure}
\centering
\includegraphics[width=14cm]{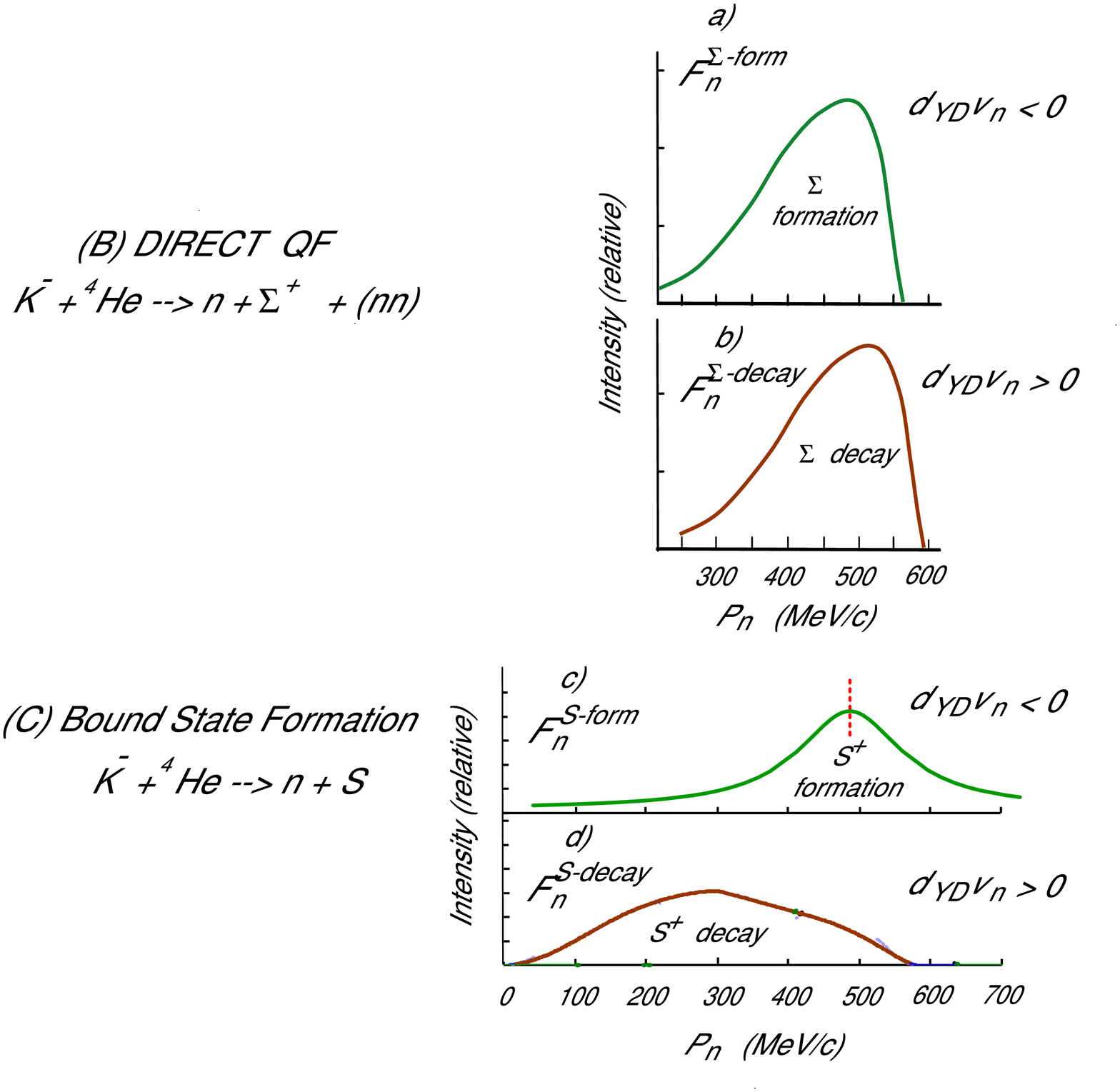}
\vspace{0cm}
\caption{\label{fig:neutron-spectrum} 
The calculated neutron spectra, a) from the QF $K^-NN$ absorption process, 
$F_n^{\Sigma-{\rm form}}$, 
b) from the decay processes of hyperons produced in the QF $K^-NN$ absorption process, $F_n^{\Sigma-{\rm decay}}$, 
and c) from the S$^+$ formation process, $F_n^{\rm S-form}$, and d) from its decay process, $F_n ^{\rm S-decay}$.
}
\end{figure}

The calculated spectra are shown in Fig.~\ref{fig:neutron-spectrum}. As shown in the preceding section, both the hyperon and nucleon momenta in the direct capture process are broadly distributed, and they are nearly identical. Then, the decay nucleons have also similar momentum spectra, because the momentum taken away by $\pi$ and $\gamma$ is relatively small. When a high-momentum pion is detected at 90 degrees with respect to the neutron emission, as realized in the KEK E470 and E549 experiments, the $\Sigma$ production and decay are the main source of the neutron spectrum above 400 MeV/c. To be more precise, we made a simple kinematical calculation, which shows that the neutron momentum from the decay of $\Sigma$ of about 480 MeV/c is boosted by about 30 MeV/c (this upward shift slightly increases to 38 MeV/c when the $\Sigma$ momentum decreases to 400 MeV/c). Thus, the neutron spectrum of the $\Sigma$ decay origin, $F_n^{\Sigma-{\rm decay}}$, moves upward by about 30 MeV/c compared with that of the absorption process, $F_n^{\Sigma-{\rm form}}$, as shown in the upper part of Fig.~\ref{fig:neutron-spectrum}. They are very similar. How can we discriminate experimentally between the two kinds of spectra?

In principle,  $F_n^{\Sigma-{\rm form}}$ and $F_n^{\Sigma-{\rm decay}}$ can be distinguished by tagging the spectrum by detecting the hyperon decay point with respect to the $K^-$ stopping point ($\vec{d}_{YD}$) because the hyperon and the nucleon are emitted almost back to back \cite{Iwasaki:03,Iwasaki:01}. When $\vec{d}_{YD}$ is opposite to the neutron direction ($\hat{v}_n$), namely, $\vec{d}_{YD} \cdot \hat{v}_n < 0$, this neutron should come from $F_n^{\Sigma-{\rm form}}$, and when $\vec{d}_{YD} \cdot \hat{v}_n > 0$, the neutron should come from $F_n^{\Sigma-{\rm decay}}$. We thus expect nearly the same neutron spectra for the two kinds of tagging. 

Now we consider possible formation of $K^-ppn$ (abbreviated as $\equiv$ S$^+$),
\begin{equation}
K^-\, + ^4{\rm He}  \rightarrow n + {\rm S}^+,
\end{equation}
in which a monoenergetic peak (narrow or broad) may appear in an inclusive neutron spectrum $F_n^{\rm S-form} (p_n)$. For $M(K^-ppn) \sim 3130$ MeV/$c^2$ the neutron momentum is $p_n \sim 480$ MeV/c. If the S$^+$ state is broad, $F_n^{\rm S-form} (p_n)$ overlaps with the continuous spectrum from  the direct capture process, and both are  mutually indistinguishable. On the other hand, the decay pattern of S$^+$,
\begin{eqnarray}
{\rm S}^+ &\rightarrow& \Sigma^+ + n + n,\\
&\rightarrow& \Sigma^0 + p + n,\\
&\rightarrow& \Sigma^- + p + p,\\
&\rightarrow& \Lambda + p + n,
\end{eqnarray}
is totally different. If a peak appears at 480 MeV/c in the inclusive neutron spectrum, the momentum of S$^+$ from the two body kinematics, is also 480 MeV/c, and both are emitted back to back. Then, the decay neutron spectrum, $F_n^{{\rm S-decay}}$, lies in a much lower momentum range than $F_n^{{\rm S-form}}$. We have calculated the decay particle spectrum from S$^+$ according to the method developed in \cite{Kienle:06}. The spectrum, as shown in Fig.~\ref{fig:neutron-spectrum}, is for the case of a shrunk nulear core in the $K^-ppn$ cluster . The tagged neutron spectrum consists of 
\begin{equation}
F_n^{\Sigma-{\rm form}} + F_n^{\rm S-form}
\end{equation} 
for $\vec{d}_{YD} \cdot \hat{v}_n < 0$, whereas
\begin{equation}
F_n^{\Sigma-{\rm decay}} + F_n^{\rm S-decay}
\end{equation} 
for $\vec{d}_{YD} \cdot \hat{v}_n > 0$. Since $F_n^{\rm S-decay}$ appears only in the low momentum region, we expect a difference between the two kinds of tagged spectrum, {\it only when a bound state S$^+$ is formed}. 

\subsection{Experimental neutron spectra}

This is the basic principle for the method to isolate the KNC component of the neutron spectrum from the QF absorption component by ``hyperon-motion tagging", as proposed in \cite{Iwasaki:03}. Of course, the decay lengths of the hyperon and S$^+$ are small and precise determination of the hyperon decay point is required. In a tagged neutron spectrum from the experiment  E471 of KEK  \cite{Iwasaki:03}, an excess fraction is observed in the 500 MeV/c region, but its statistical significance was low. It is extremely interesting to apply this method to new experimental data from E547 of KEK \cite{Iwasaki:06} and also from AMADEUS \cite{AMADEUS}.

\section{Invariant masses of $\Lambda$-$p$}

\subsection{Partial invariant masses and quasi-invariant masses}

The invariant mass of a particle pair ($x_1 - x_2$) may indicate presence of a parent resonance state $X$ ($\rightarrow x_1 + x_2$), for instance, $K^-pp$ ($\rightarrow \Lambda + p$). However, when such a resonance state is embedded in a nuclear medium, namely, in an ``off-shell" state, the validity of relativistic invariance is lost, and thus, the ``invariant mass" loses its original genuine meaning, though an ``invariant mass" can be experimentally reconstructed. The implications of such invariant masses of off-shell particles, called {\it quasi-invariant masses (QIM)}, are fully discussed in the paper \cite{Yamazaki:99} to clarify the prevailing misunderstanding and naive interpretation of  a ``shifted invariant mass" in terms of  a ``medium-modified hadron mass". 

There are two important effects. First, the invariant mass reconstructed from decay particles from a bound state of $X$ is not the mass of $X$ at all. The {\it QIM} of $X$ is given by $M_{\rm inv} (X) = \sqrt{E_X^2/c^4 - P_X^2/c^2}$, with $P_X$ being the internal momentum. Since $P_X$ is not a discrete quantity, the {\it QIM} of $X$ distributes broadly below the bound-state energy. The second effect is shift and broadening of {\it QIM} due to collisions with surrounding nucleons in which the decay process is involved, such as in $XN \rightarrow x_1 + x_2 + N'$. This latter effect (collisional shift and broadening) was shown to be significant in the decay process $\rho \rightarrow e^+ + e^-$, when the $\rho$ meson is embedded in a dense nuclear medium \cite{Yamazaki:99}. 

In the OT Ansatz that $M([K^-pp]_{\rm atom}) = M_0 - B_{\alpha}/c^2$, eq.(\ref{eq:OT-MKNN}), they assume the $[K^-pp]_{\rm atom}$ state is on-shell, without coupling to the remaining system. Thus, the invariant mass is discrete, nothing but $M([K^-pp]_{\rm atom})$. On the other hand, in our view,  the $[K^-pp]_{\rm atom}$ subsystem cannot exist without binding to $[nn]$ in $[K^-\,^4$He]$_{\rm atom}$, and is regarded as a bound state with a large internal momentum $P$. Thus, its quasi-invariant mass, {\it QIM}, 
\begin{equation}
M_{\rm inv} ([K^-pp]_{\rm atom}) = \sqrt{(M_0 - B_{\alpha}/c^2)^2 - P^2/c^2} \label{eq:Minv-KNN},
\end{equation}
has a distribution according to the distribution in $P$. Realistic calculations of $M_{\rm inv}(\Lambda p)$ will be presented later.

\subsection{Dalitz domains}
It is convenient to discuss various reactions and decay processes in terms of Dalitz domains. The three-body decay of a $\bar{K}NN$ system was discussed by Kienle {\it et al.} \cite{Kienle:06}, in which the dynamical effect of the three-body correlation arising from possible shrinkage of the $\bar{K}$ bound system as well as from the spin and parity involved was predicted. In general, the kinematics of a three-body decay process in (\ref{eq:AY-KNN-1b}, \ref{eq:AY-KNN-1a}) can be expressed in a Dalitz plot with squared partial invariant masses, $m(Y p)^2$ and $m((nn) Y)^2$, which are related to the c.m. energies, $E_{(nn)}$ and $E_p$, respectively, as
\begin{eqnarray} 
m(Y p)^2&=&M^2 + M_{(nn)}^2 - 2\,M E_{(nn)} /c^2,\label{eq:m-E-1}\\
m((nn) Y)^2&=& M^2 + M_{p}^2 - 2\,M E_p /c^2,\label{eq:m-E-2}
\end{eqnarray}
where $M = m_K + M(^4{\rm He})$ = 4221 MeV/$c^2$.

\begin{figure}
\centering
\includegraphics[width=10cm]{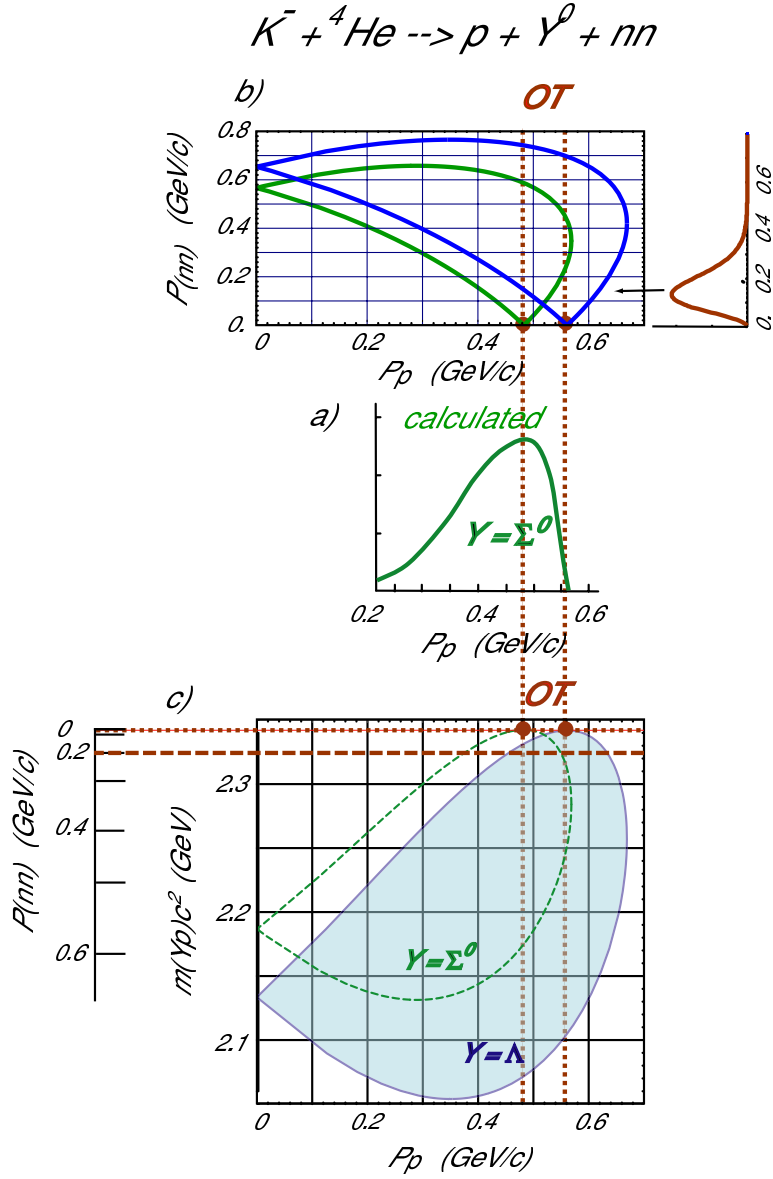}
\vspace{0cm}
\caption{\label{fig:P_p} 
 The proton momentum spectrum in the case of $K^- + ^4$He $\rightarrow p + Y^0 + (nn)$ reaction at rest.  a) Calculated distribution of $P_p$ with the realistic wave function of $^4$He. 
  b) Dalitz domains in the $P_{(nn)}$ vs $P_p$ variables for $Y^0 = \Lambda$ (blue curve) and $Y^0 = \Sigma^0$ (green curve). The OT points corresponding to $P_{(nn)} = 0$ are shown by brown dots, whereas the realistic distribution of $P_{(nn)}$ is shown on the right-hand side. c) Dalitz domains  in the $m(Yp)c^2$ vs $P_p$. The Oset-Toki prediction corresponds to $M_{\rm inv}(\Lambda p)$ = 2342 MeV, designated by a closed circle and a dotted horizontal line, whereas the $M_{\rm inv}(\Lambda p)$ expected from the recoil momentum of $P_{(nn)} \sim 0.2$ GeV/$c$ is shown by a horizontal dashed line.
 }
\end{figure}

Fig.~\ref{fig:P_p} (B) shows Dalitz domains with the momentum variables, $P_p$ and $P_{(nn)}$, which are converted according to the relations, (\ref{eq:m-E-1}, \ref{eq:m-E-2}). 
OT claim that the recoil momentum is zero, namely, $P_{(nn)} = 0$, leading to two OT points, shown in the figure. They correspond to the two discrete lines in $P_p$, as given in eq.(\ref{eq:OT-value-Sigma0}, \ref{eq:OT-value-Lambda}). On the other hand, the realistic distribution of $P_{(nn)}$ given in Fig.~\ref{fig:d-momentum} is shown on the right-hand side of the figure. This gives a guide line as to how broad the spectrum of $P_p$ should be.

 Fig.~\ref{fig:P_p} (C) shows another presentation of the Dalitz domain in a $P_p$-$m(Yp) c^2$ plane, where the y axis in Fig.~\ref{fig:P_p} (A) is converted from $P_{(nn)}$ to $m(Yp) c^2$, the partial invariant mass of $Y$ and $p$. The OT points are also shown by two dots. The invariant mass, $m(Y p)$, in the decay process of OT (\ref{eq:KNN-OT-1}, \ref{eq:KNN-OT-2}) with the fictitious mass (\ref{eq:OT-MKNN})  is shown by a dotted horizontal line. 
 

\subsection{Calculations of $M_{\rm inv} (\Lambda p)$}

\begin{figure}
\centering
\includegraphics[width=16cm]{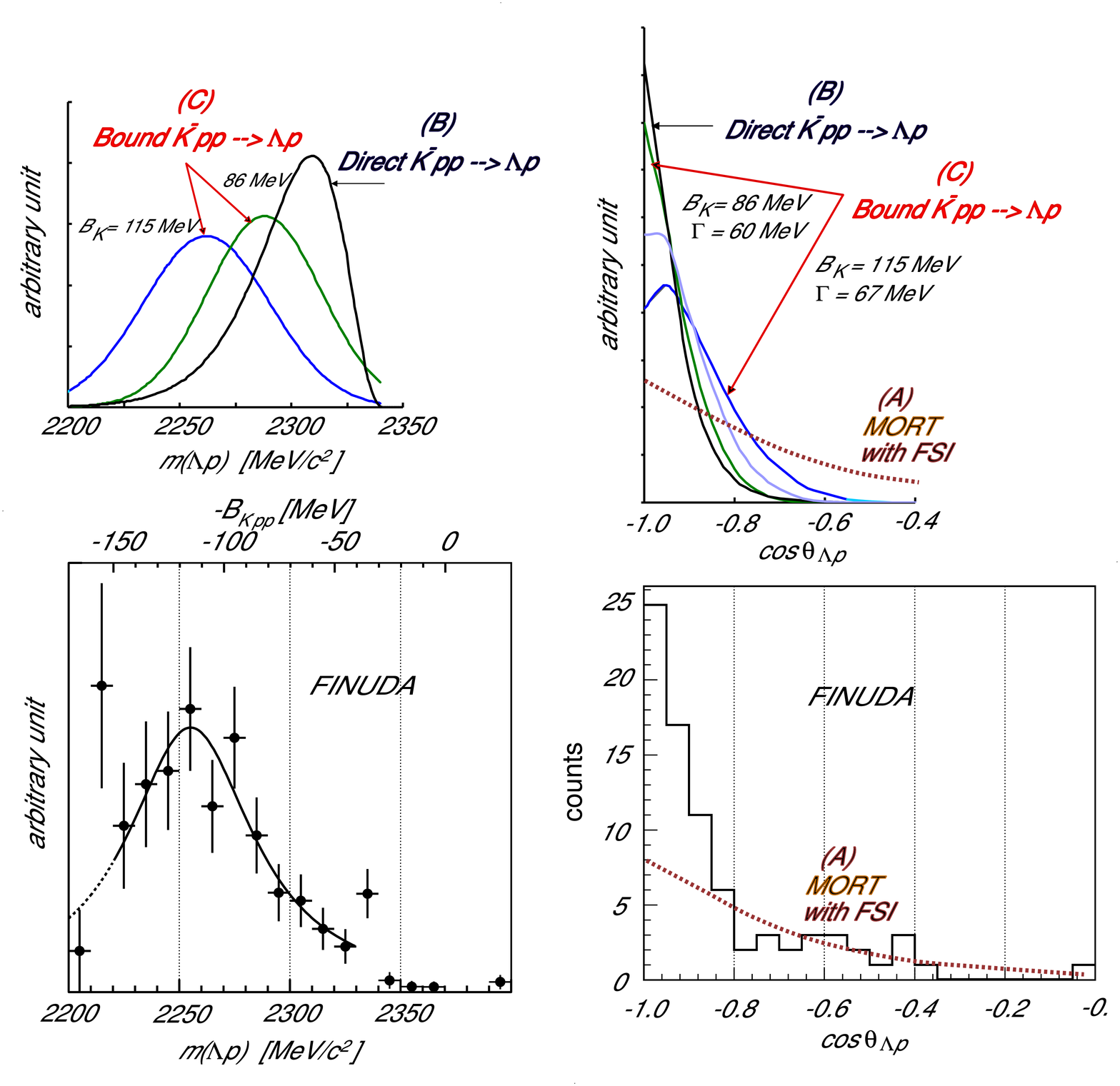}\vspace{0.cm}
\caption{\label{fig:Lambda-p} 
(Left-Upper) The calculated $M_{\rm inv}(\Lambda p)$ distribution in the $K^- + ^4$He reaction for the QF $K^-pp \rightarrow \Lambda + p$ decay process. Additional two curves are obtained when a KNC ($K^-pp$) is formed with a binding energy of 86 MeV and 115 MeV. They are compared with the FINUDA spectrum (Left-Lower). The Oset-Toki prediction corresponds to $M_{\rm inv}(\Lambda p)$ = 2342 MeV.
(Right-Upper)
The calculated angular correlation of $\Lambda - p$ pair in $K^- + ^4$He in the case of a QF $K^-pp \rightarrow \Lambda + p$ process and in the cases of KNC formation. The Oset-Toki prediction corresponds to cos\,$\theta_{\Lambda p} \sim -1.0$, whereas MORT with FSI interaction predict a broader distribution for $^7$Li, as shown by the dotted curve. The FINUDA histogram for $^6$Li, $^7$Li and $^{12}$C \cite{FINUDA-Fujioka}, as shown below, is in good agreement with our curves, whereas it seriously contradicts  the MORT prediction with FSI.}
\end{figure}

We perform realistic calculations of partial invariant masses composed of $\Lambda$ and $p$ in the  absorption processes, (B) and (C). Namely, we calculate $\vec{P}_\Lambda$ and $\vec{P}_p$ realistically and then reconstruct their invariant masses. 
The calculated invariant mass $M_{\rm inv}(\Lambda p)$ is shown in Fig. \ref{fig:Lambda-p} (Left).  It distributes around 2310 MeV/$c^2$ with a FWHM of about 50 MeV/$c^2$. It is equivalent to $P_{(nn)}$, which also distributes between 200 and 300 MeV/$c$. The calculated $\Lambda$-$p$ angular correlation is shown in Fig.~\ref{fig:Lambda-p} (Right).  

Unfortunately, there is no corresponding experimental data available in the $K^-$ + $^4$He absorption, but we can compare with the $M_{\rm inv}(\Lambda p)$ spectrum of FINUDA \cite{FINUDA-PRL}, though the targets are somewhat different ($^6$Li, $^7$Li and $^{12}$C). The FINUDA spectrum reveals a large bump at $Mc^2 \sim 2260$ MeV with a small excursion in one bin at $\sim$ 2340 MeV. So, we are tempted to examine what we expect when a nuclear bound state is formed. 

The KNC process (C) is
\begin{equation}
K^- \,^7{\rm Li} \rightarrow [K^-pp]_{\rm nucl} + t + 2n
                          \rightarrow \Lambda + p  + t + 2n, 
\end{equation}
where $[t + 2n]$ represents also an ensemble of $[d+3n]$ and $[p+4n]$. The KNC process (\ref{eq:KNC-AY-1}) releases a large energy because of a large \kbar~binding, and thus, one might wonder that,  if  $[K^-pp]_{\rm nucl}$ with a binding energy of $B_K = 100$ MeV is formed with a two-body decay partner like $^5$H, the energy-momentum conservation would give a momentum of about 600 MeV/$c$ to $[K^-pp]_{\rm nucl}$. This view is not consistent with the fact that the momentum of the $\Lambda-p$ system actually distributes around $200-300$ MeV/$c$ \cite{FINUDA-Fujioka}, and that the $\Lambda$-$p$ pair shows a back-to-back correlation. On the other hand, when the recoil system is composed of free $[t + 2n]$ particles, the above constraint is relaxed. Namely, the $P_{(\Lambda p)}$ becomes 200 - 300 MeV/$c$ and the $\Lambda$-$p$ angular correlation is nearly back-to-back. This is in good agreement with the observation. 

The predicted binding energy of $K^-pp$ KNC is 48 MeV and its mass is 2322 MeV/$c^2$ \cite{Yamazaki:02}. In the invariant-mass spectrum this bound state peak nearly overlaps with the peak arising from the Direct QF absorption process (B). The FINUDA spectrum, however, shows no such peak.    
We also present invariant-mass spectra for $\bar{K}$ binding energies of 86 MeV and 115 MeV in Fig.~\ref{fig:Lambda-p} (Left) together with the FINUDA data. The observed spectrum seems to indicate the case of $B_K$ = 115 MeV. The reason why the direct $K^- pp$ absorption peak is absent in the FINUDA spectrum is not understood. The claimed bump around 2260 MeV/$c^2$ will be studied furthermore in a forthcoming improved experiment.    

\begin{figure}
\centering
\includegraphics[width=7cm]{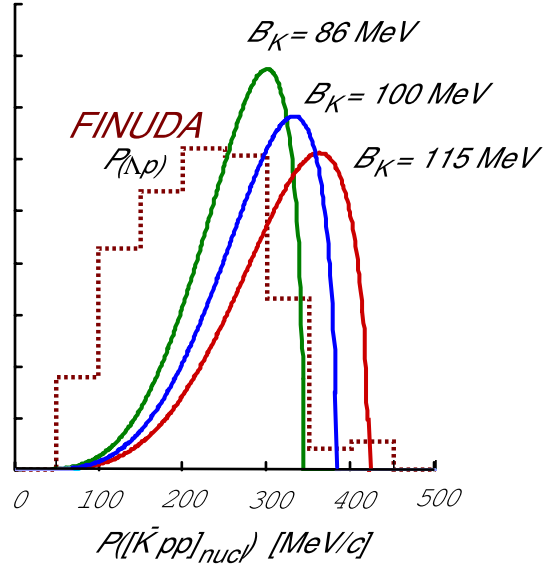}
\vspace{0.cm}
\caption{\label{fig:P(Kpp)-FINUDA} 
The calculated $P([K^-pp]_{\rm nucl})$ distribution in $K^- + ^4$He in the cases of KNC formation. The experimental data, $\vec{P}_{(\Lambda p)} = \vec{P}_{\Lambda} + \vec{P}_p$, of FINUDA \cite{FINUDA-Fujioka} is plotted as a dotted histogram. }
\end{figure}

The calculated momentum distributions of $P([K^-pp]_{\rm nucl})$ are shown in Fig. \ref{fig:P(Kpp)-FINUDA}. In the same figure we show an observed momentum distribution of $\vec{P}_{(\Lambda p)} = \vec{P}_{\Lambda} + \vec{P}_p$ in the $\Lambda$-$p$ pair events in the FINUDA experiment \cite{FINUDA-Fujioka}, which indicates the presence of a large recoil momentum (200 - 300 MeV/$c$), consistent with the calculated shapes. The calculated $\Lambda$-$p$ angular correlation still show a back-to-back feature, as shown in Fig.~\ref{fig:Lambda-p} (Right), in good agreement with the observed angular correlation.

\subsection{MORT's interpretation of FINUDA $M_{\rm inv} (\Lambda p)$}

MORT \cite{MORT} calculated the invariant-mass spectrum, $M_{\rm inv} (\Lambda p)$, taking the $K^-pp \rightarrow \Lambda + p$ process into account in
the $K^- + ^7{\rm Li}$ (and other) absorption. They used an appropriate $K^-$ wavefunction for the capture process, which would {\it inevitably indicate that the nucleons involved in the $K^-$ capture have Fermi momenta around 200 MeV/$c$ so that no discrete line in $P_p$ nor $P_\Lambda$ would be anticipated, contrary to what OT incorrectly insisted.} On the other hand, the widely distributed momentum distributions result in rather a sharp peak in an invariant-mass spectrum, since the Fermi motion effects are more or less cancelled out in $M_{\rm inv} (\Lambda p)$. In the following we discuss this problem. 

\begin{figure}
\centering
\includegraphics[width=12cm]{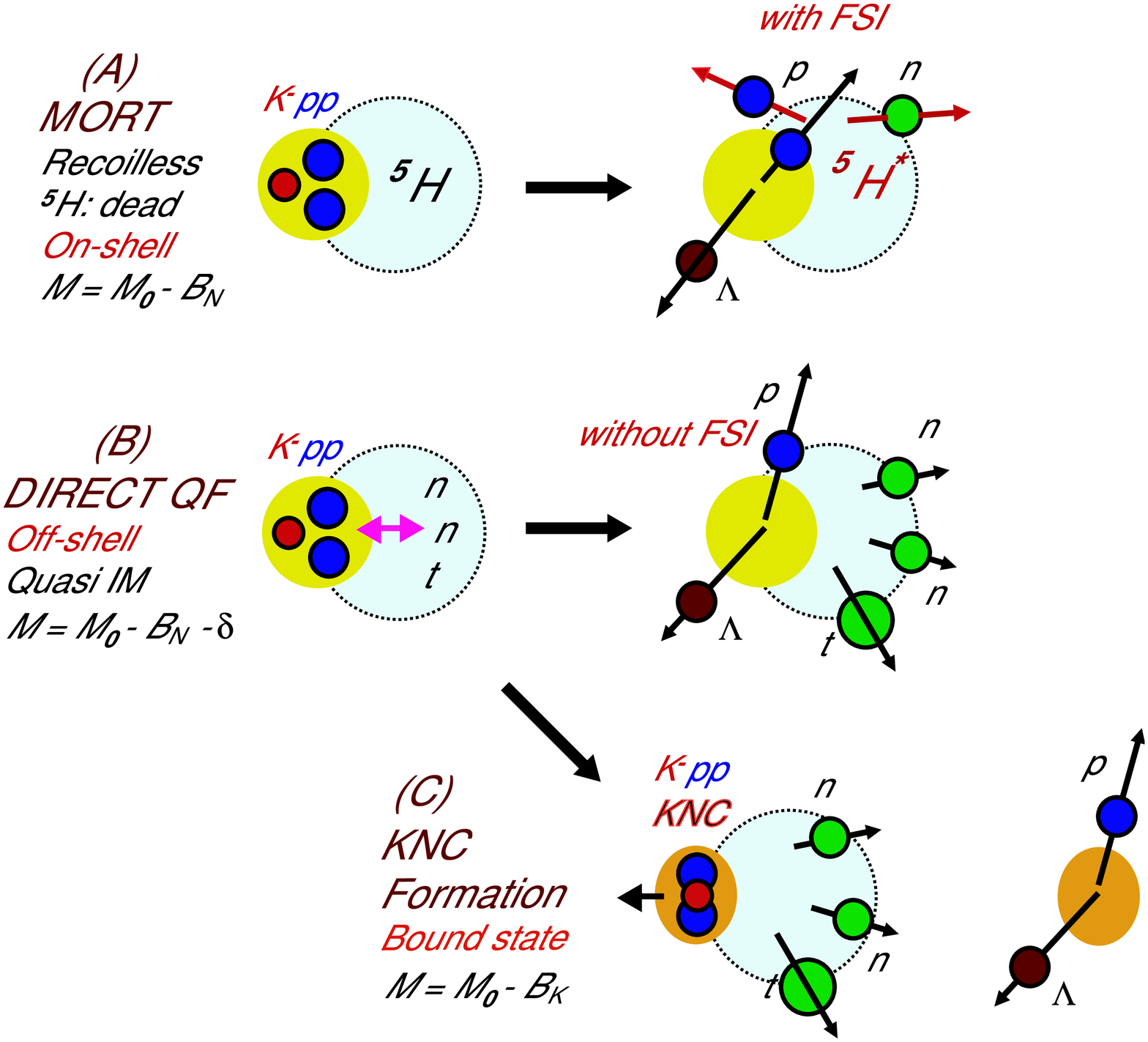}
\vspace{0cm}
\caption{\label{fig:Kabs-mechanism} 
Three different $K^-$ absorption mechanisms (in the case of $^7$Li as an example).  (Upper) OT/MORT mechanism. (Middle) Direct QF $K^- pp \rightarrow \Lambda + p$. (Lower) Formation of a nuclear bound state $K^- pp$.}
\end{figure}

MORT calculate the following process,
\begin{eqnarray}
[K^- \,^7{\rm Li}]_{\rm atom} &=& [K^-pp]_{\rm atom} + {^5{\rm H}}\\
&&[K^-pp]_{\rm atom} \rightarrow p + \Lambda,
\end{eqnarray}\label{eq:7Li-5H-2body-absorption}
where $[K^-pp]_{\rm atom}$ is an on-mass-shell object consisting of grouping of $pp$ captured by the atomic $K^-$, which inherits the nuclear binding energy of $^7$Li ($B_N \sim 30$ MeV). This on-mass-shell treatment leads to an invariant mass $M_{\rm inv}(\Lambda p)$:
\begin{equation}
M_{\rm inv}(\Lambda p) = M([K^-pp]_{\rm atom})
= M_0 - B_N
 = 2340~{\rm MeV}/c^2.
\end{equation}
In this scheme the residual nucleus, $^5$H, has a mass nearly equal to that of $t + 2n$, and can receive no momentum from this two-body decay process, since the energy conservation does not allow the $^5$H to be energetic. Apparently, this absorption process contradicts the FINUDA spectrum where the 2340 MeV peak is only a very small fraction (at most $\sim 6~\%$) of $\Lambda p$ pairs. To account for the majority of $\Lambda p$ events whose invariant mass lies far below this line MORT invoke a FSI effect in which most of the emitted $\Lambda$ and $p$ undergo quasi-elastic scattering in the assumed residual nucleus $^5$H (see Fig.~\ref{fig:Kabs-mechanism} (A)).  

\begin{figure}
\centering
\includegraphics[width=12cm]{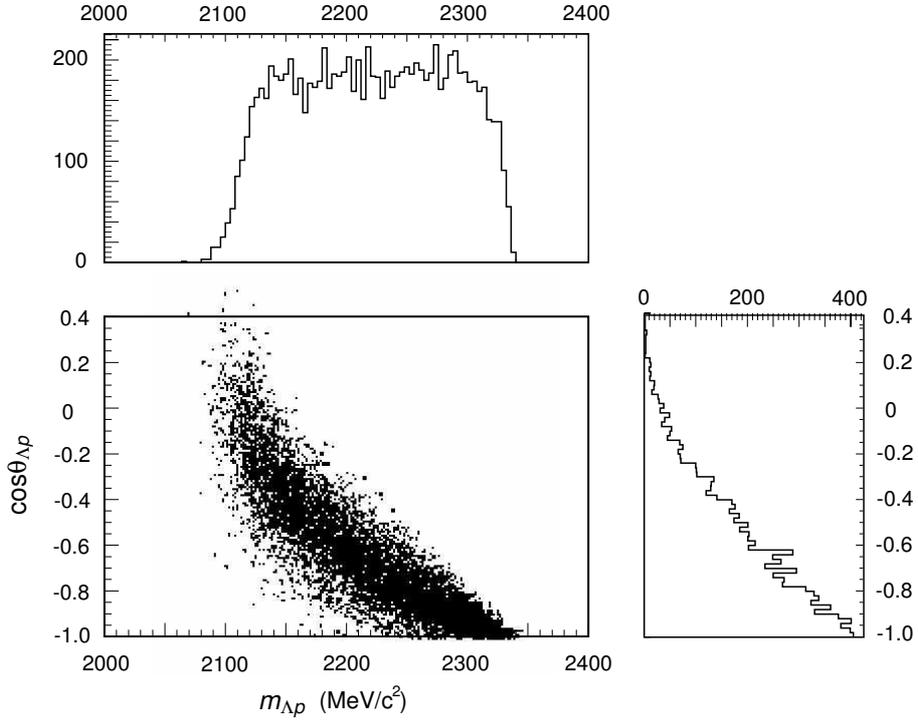}
\vspace{0.cm}
\caption{\label{fig:M-theta-correlation} 
Correlation between $M_{\rm inv} (\Lambda p)$ and cos$\theta_{\Lambda p}$ for simulated events in $K^- \,^7$Li after FSI. They are distributed along a unique locus. From Hayano \cite{Hayano:06}.}
\end{figure}

In general, such a FSI effect causes substantial shift and broadening of the original $M_{\rm inv}(\Lambda p)$ to a lower-mass side, producing a nearly flat $M_{\rm inv}(\Lambda p)$ distribution. At the same time, it destroys the original back-to-back angular correlation between $\Lambda$ and $p$. The ${\rm cos\,}\theta_{\Lambda p}$ distribution of MORT, as reproduced in Fig.~\ref{fig:Lambda-p} (Right), is much broader than those in the QF and KNC cases, and is in serious disagreement with the FINUDA observation, that is, the observed angular correlation is sharp without any kinematical cut and is not affected by the relevant detection efficiency \cite{FINUDA-PRL,FINUDA-Fujioka}. Nevertheless, MORT selected a back-to-back fraction ($-1 < {\rm cos\,}\theta_{\Lambda p} <-0.8$) of their broad angular correlation and reconstructed $M_{\rm inv}(\Lambda p)$ under this cut. {\it This artificial procedure inevitably produces a peak-like fake structure, by which MORT claim to have accounted for the FINUDA peak}. However, since $M_{\rm inv}(\Lambda p)$ and ${\rm cos\,}\theta_{\Lambda p}$ are uniquely correlated, namely, with the increase of the FSI effect (corresponding to large-angle scattering) the $M_{\rm inv}(\Lambda p)$ decreases and the ${\rm cos\,}\theta_{\Lambda p}$ is broadened, as shown in Fig.~\ref{fig:M-theta-correlation}, {\it their cut on a smaller ${\rm cos\,}\theta_{\Lambda p}$ (back-to-back correlation) is equivalent to artificially selecting a larger $M_{\rm inv}(\Lambda p)$ region, thereby  introducing a ``self-produced" fake}. On the other hand, the peak like structure at 2260 MeV/$c^2$ in the experimental spectrum of $M_{\rm inv}(\Lambda p)$ in FINUDA is hardly affected by their angular correlation cut, since the observed angular correlation is nearly back-to-back.  

\begin{figure}
\centering
\includegraphics[width=12cm]{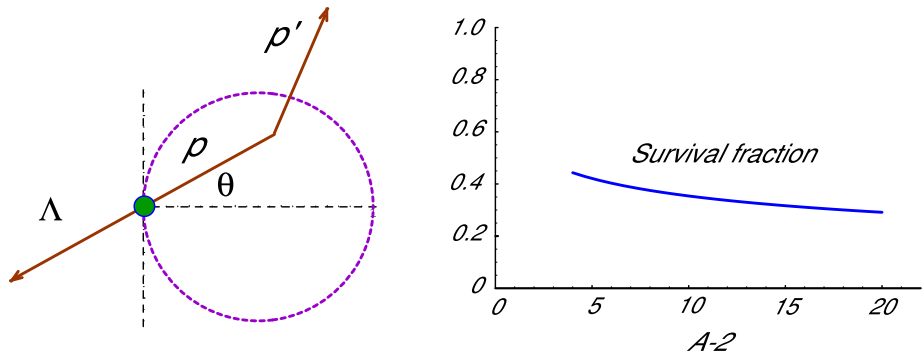}
\vspace{0.cm}
\caption{\label{fig:FSI} 
(Left) Geometry of FSI of a proton, produced at the nuclear surface of a residual nucleus of mass number $A-2$ and emitted with an angle $\theta$. (Right) Estimated survival probability of the $M_{\rm inv}$ peak at $m_K + 2 M_N$.}
\end{figure}

Furthermore, MORT ought to postulate a very small ``survival" fraction ($\sim 6 \%$) for the FSI effect to explain the very small intensity of the ``surviving" 2340 MeV/$c^2$ peak in the FINUDA observation. We estimated the survival fraction by the following simple procedure. Since the invariant mass, $M_{\rm inv} (\Lambda p)$, is changed by a single scattering of either the proton or $\Lambda$, the survival probability of the original peak at $M_{\rm inv} = m_K + 2 \, M_N$ is given by
\begin{equation}
S =< {\rm exp}[-\frac{2 R\, {\rm cos} \theta}{L}]>,
\end{equation}
where $R$ is the nuclear radius of the residual nucleus $[A-2]$, and $L = (\sigma \rho)^{-1} = 1.96$ fm  (with $\sigma = 30$ mb and $\rho = 0.17$ fm$^{-3}$). In this simple treatment we assumed that the proton, produced at the nuclear surface and emitted with an emission angle $\theta$, is scattered by a nucleon in the residual nucleus with the normal nuclear density ($\rho_0$) (see Fig.\ref{fig:FSI}). 

The above formula gives the survival probability versus $A-2$, as shown in Fig.~\ref{fig:FSI}. It is about 44\% in  $K^- \,^6$Li,  42 \% in $K^- \,^7$Li and 35 \% in $K^- \,^{12}$C, much larger than 6 \%. When the residual nucleus has a diffused density distribution, smaller than 0.17 fm$^{-3}$, presumably as in the case of $^5$H, which has no ground state, but a resonance state at 1.7 MeV above the $t+n+n$ mass with a width of 1.9 MeV \cite{H5}, the survival fraction becomes larger. Thus, the above estimate is regarded as a lower limit. 

MORT seem to say that a large fraction (say, $\sim 70$ \%) of $\Lambda$-$p$ in the $K^-$ absorption process does not correspond to the ground state of $^5$H, and thus, does not contribute to the 2340 MeV/$c^2$ peak. Then, what kind of invariant mass does this fraction produce? This fraction is what we treat in our Direct QF process (B) with residual fragments of $t + 2n$, which carry appreciable but limited  momenta.  In our QF process there is ``no ground state", and the particle emission takes place in continuum (see Fig.~\ref{fig:Kabs-mechanism} (B)) from the beginning. Nevertheless, $M_{\rm inv}(\Lambda p)$ is concentrated in the region near the threshold ($\sim 2310$ MeV/$c^2$). If the ``70 \% non-ground-state fraction" of MORT could produce invariant masses much lower than this limit, the energy and momentum conservation would require that the momenta of the residual fragments should be very much beyond their internal momenta. Of course, this is unlikely. 

\section{Comparison between $K^-$ + $^6$Li and $K^-$ + $^4$He}

\begin{figure}
\centering
\includegraphics[width=11cm]{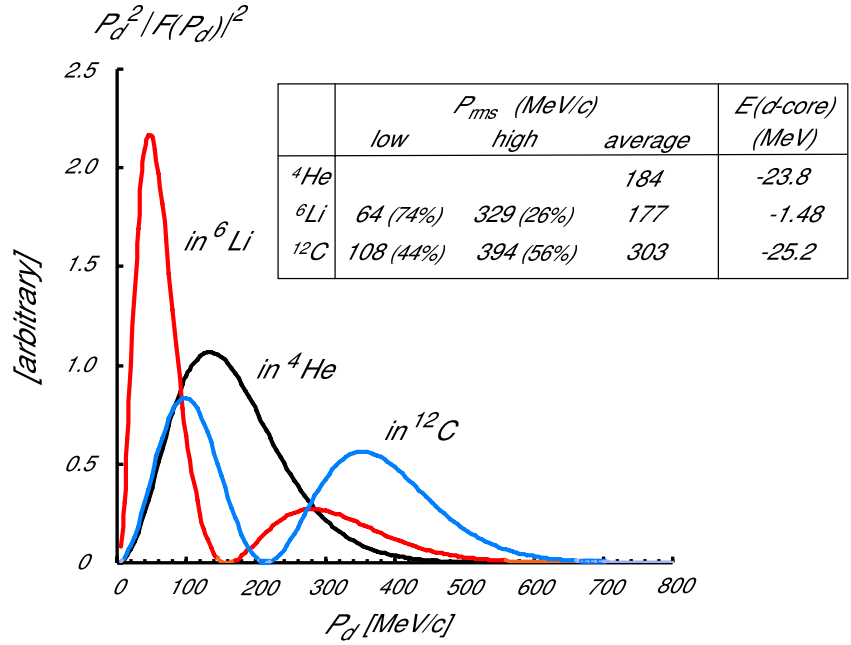}
\vspace{-0cm}
\caption{\label{fig:d-momentum} 
The calculated momentum distributions of quasi-$d$ in typical light nuclei, $^4$He, $^6$Li and $^{12}$C. The binding energies of quasi-$d$ in these nuclei and the rms momenta are shown in the inset. }
\end{figure}

Since FINUDA reported an inclusive proton spectrum from $K^-$ + $^6$Li \cite{FINUDA-6Li}, it is worthwhile to study the case of $^6$Li realistically. We have shown in Fig.~\ref{fig:d-momentum} that $^6$Li is only the one exceptional case of an nearly isolated $d$ cluster, namely, 
\begin{equation}
^6{\rm Li} \approx d + \alpha,
\end{equation}
with a very small $d-\alpha$ binding energy of 1.48 MeV and a very small rms momentum of 64 MeV/$c$. Thus, the $K^-$ absorption is expected to occur either in $^4$He or $d$, namely,
\begin{eqnarray}
{\rm (K-}\alpha)~~ &&K^- +~ ^6{\rm Li} \rightarrow d + p + X,\\
{\rm (K-} d)~~ &&K^- +~ ^6{\rm Li} \rightarrow p + \Sigma^- +~ ^4{\rm He},
\end{eqnarray}
respectively. It is worthwhile to examine whether or not the same kind of peak $X = $S$^0$(3115), as once reported at $P_p \sim 500$ MeV/$c$ in the KEK E471 experiment \cite{Suzuki:04}, may persist.

\begin{figure}
\begin{center}
\vspace{0cm}
\includegraphics[width=12cm]{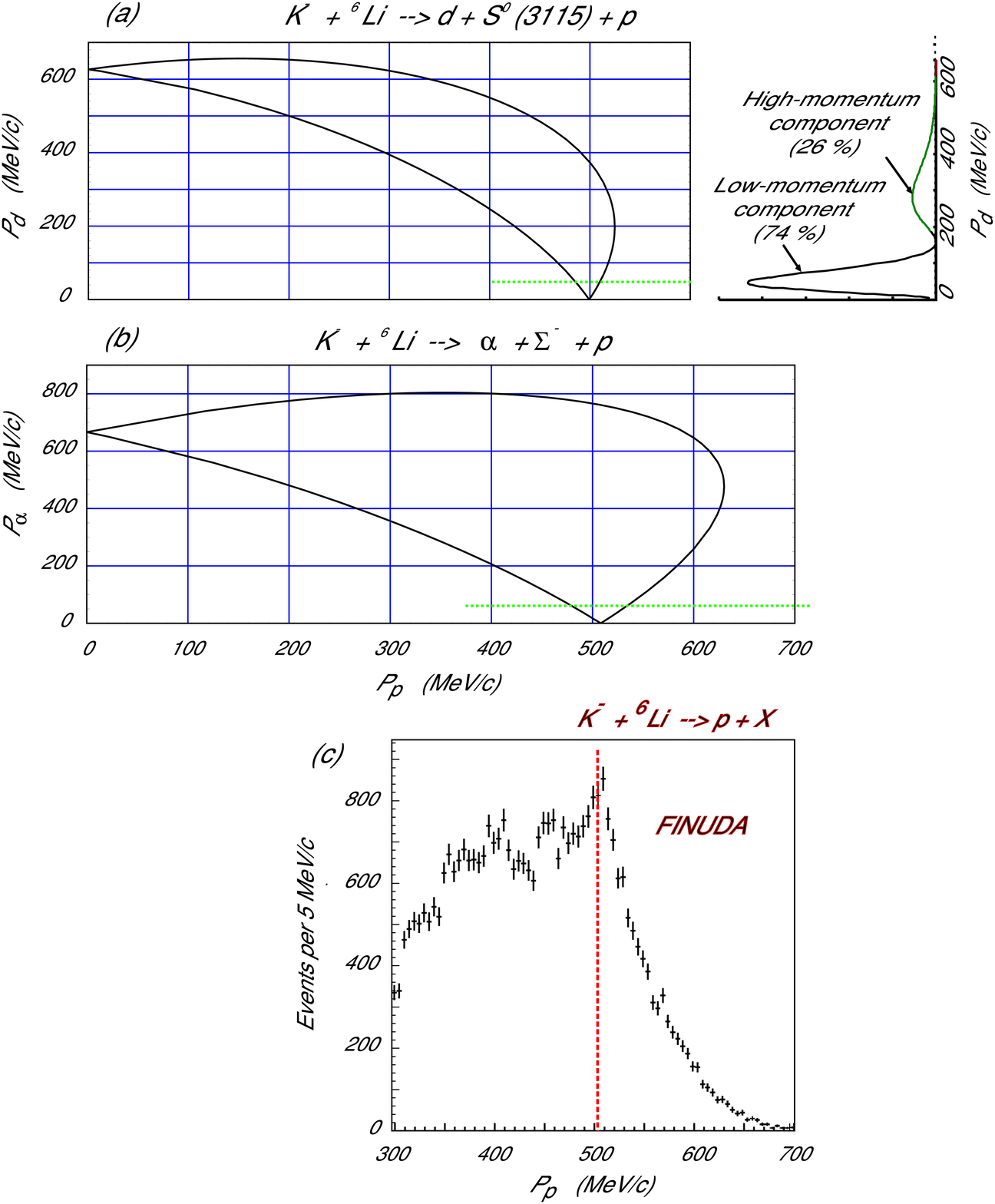}
\vspace{0cm}
\end{center}
\caption{\label{fig:FINUDA-6Li}
(a) Dalitz domain of $P_d$ vs $P_p$ in $K^- + {^6{\rm Li}} \rightarrow d + {\rm S}^0 (3115) + p$. 
(b) Dalitz domain of $P_{\alpha}$ vs $P_p$ in $K^- + {^6{\rm Li}} \rightarrow \alpha + \Sigma^-  + p$. (c) Inclusive spectrum of $P_p$ in the $K^-$ + $^6$Li reaction observed by FINUDA \cite{FINUDA-6Li}.
}
\end{figure}

\begin{figure}
\begin{center}
\vspace{0cm}
\includegraphics[width=14cm]{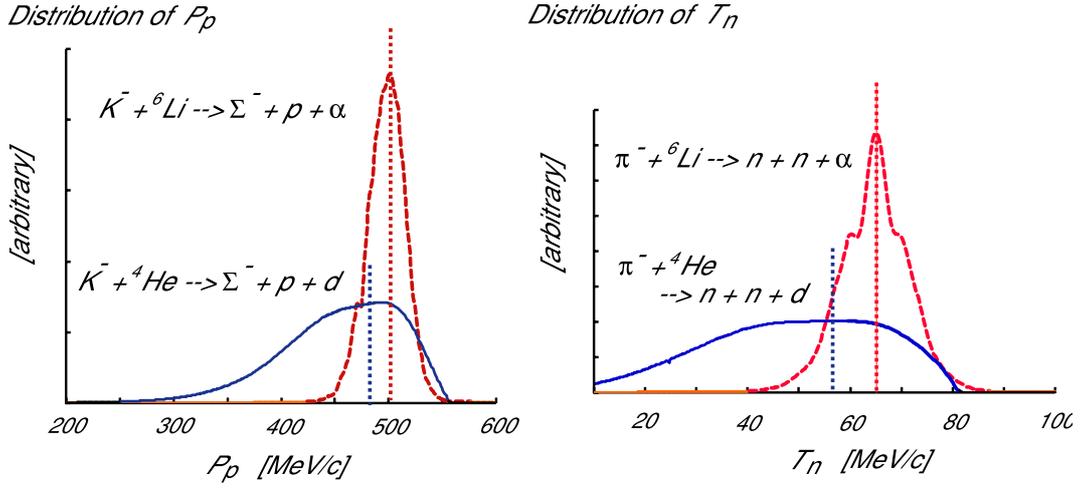}\vspace{0cm}
\end{center}
\caption{\label{fig:P_p-6Li-4He} (Left) Calculated proton momentum spectra from $K^- + {^4 {\rm He}} \rightarrow \Sigma^- + p + d$ (solid blue) and $K^- + {^6 {\rm Li}} \rightarrow \Sigma^- + p + {^4 {\rm He}}$ (broken red). (Right) Calculated neutron energy spectra from $\pi^- + ^4 {\rm He} \rightarrow n + n  + d$ (solid blue) and $\pi^- + ^6 {\rm Li} \rightarrow n + n + ^4 {\rm He}$ (broken red). The positions of the momenta in the ``recoilless case" are denoted by dotted vertical lines.}
\end{figure}

Fig.~\ref{fig:FINUDA-6Li} (a) shows a Dalitz type plot in a plane $P_d$ vs $P_p$ for a three-body decay of the absorption (K-$\alpha$) with $d$ as a spectator. The calculated rms momentum $P_d \sim 50$ MeV/$c$ is indicated by a horizontal dotted line. Hence, we recognize that a signature of S$^0$(3115) may appear at the same momentum without much broadening in an inclusive proton spectrum from $^6$Li. The other absorption process (K-$d$) is a two-body production of $p$ and $\Sigma^-$ with a small recoil momentum of the spectator $\alpha$ particle. Also in this case, the discreteness of the $p$ and $\Lambda$ particles is not washed away, and a peak in an inclusive proton momentum spectrum is expected at 508 MeV/$c$. This expectation is proved by a realistic calculation of proton spectra after $K^-$ capture by $^4$He and $^6$Li, as shown in Fig.~\ref{fig:P_p-6Li-4He} (Left).  

In fact, FINUDA observed a peak around 500 MeV/$c$ from $^6$Li but not from $^{12}$C \cite{FINUDA-6Li}, as shown in Fig.~\ref{fig:FINUDA-6Li} (c). In the latter case, the quasi-deuteron has a large momentum, as shown in Fig.~\ref{fig:d-momentum}, so that no discrete line is expected. Thus, the observed bump around  500 MeV/$c$ in the FINUDA $^6$Li spectrum may originate from these two independent origins. It is pointed out that the peak region has a strong correlation with a high-momentum $\pi^-$, which comes from $\Sigma^-$ decay. This fact is compatible with the process (K-$d$) \cite{FINUDA-6Li}. On the other hand, the $\pi^-$ from the process (K-$\alpha$) should have lower momentum below 200 MeV/$c$ and no strong correlation with the 500 MeV/$c$ proton. Then, FINUDA concluded that the 500 MeV/$c$ peak mainly originates from the (K-$d$) $K^- + d$ reaction, and thus casted a serious question about the previous KEK result \cite{Suzuki:04}. On the other hand, OT's Ansatz leads to a conclusion that the 500 MeV/$c$ proton peak comes from both $d$ and $^4$He in $^6$Li. As discussed repeatedly here, the $^4$He in $^6$Li cannot be a source for a discrete proton of $K^- + d$ origin. Thus, the KEK experiment was incompatible with the FINUDA data. In fact, the recent renewed experiment at KEK does not show such a discrete peak, and thus removed this discrepancy.  

The FINUDA paper \cite{FINUDA-6Li} says that the quasi-deuteron like character persists in $^4$He itself and other nuclei. This is a misleading claim. What matters here is how much momentum transfer is involved in a ``deuteron-cluster" reaction. In a well known $\pi^-$ capture by $^4$He, a quasi-deuteron type reaction,
\begin{equation}
\pi^- + {^4 \rm He} \rightarrow n + n + d,
\end{equation}
takes place, but this process is not recoilless, and washes away the mono-energetic character of two emitted neutrons. The calculated neutron energy spectra from $\pi^-$ capture in $^4$He and $^6$Li, shown in Fig.~\ref{fig:P_p-6Li-4He} (Right), prove this expectation. Indeed, there is no peak at 56 MeV in an observed neutron energy spectrum and only one distinct  peak appeared at 90 MeV, which comes from $t + n$ final state \cite{Cernigoi:81}. On the other hand, the $\pi^-$ absorption reaction in $^6$Li produced a 67 MeV monoenergetic neutron \cite{Cernigoi:86}, consistent with our view.  

\begin{figure}
\begin{center}
\vspace{0cm}
\includegraphics[width=8cm]{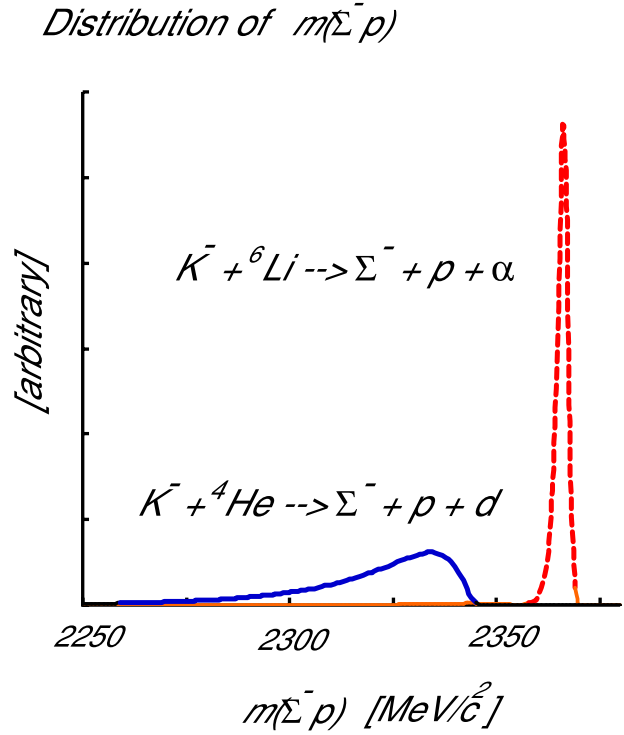}
\vspace{0cm}
\end{center}
\caption{\label{fig:m(Sp)-6Li-4He} Calculated invariant-mass spectra of $M_{\rm inv} (\Sigma^- p)$ from $K^-$ absorption in $^4$He (solid blue) and $^6$Li (broken red) assuming the Direct QF process (B).}
\end{figure}

Figure \ref{fig:m(Sp)-6Li-4He} shows calculated invariant-mass spectra of $\Sigma^- -p$ after $K^-$ capture at rest by $^4$He and $^6$Li in the case of Direct QF process, though there is no experimental data yet. The peak is much sharper in the case of $^6$Li.

\section{Summary}

We have calculated the emission and (partial) invariant-mass spectra of $N$ and $Y$ in the at-rest capture of $K^-$ in $^4$He according to the direct quasi-free process (B). The momentum spectra of $N$ and $Y$ are very broad (around 200 MeV/c), reflecting the internal momenta of the relevant nucleons in the target nucleus, whereas the invariant-mass, $M_{\rm inv} (YN)$, is concentrated in a narrow region close to the upper kinematical limit, leaving the residual $N+N$ system with a momentum around 200 MeV/c.

\begin{figure}
\centering
\includegraphics[width=14cm]{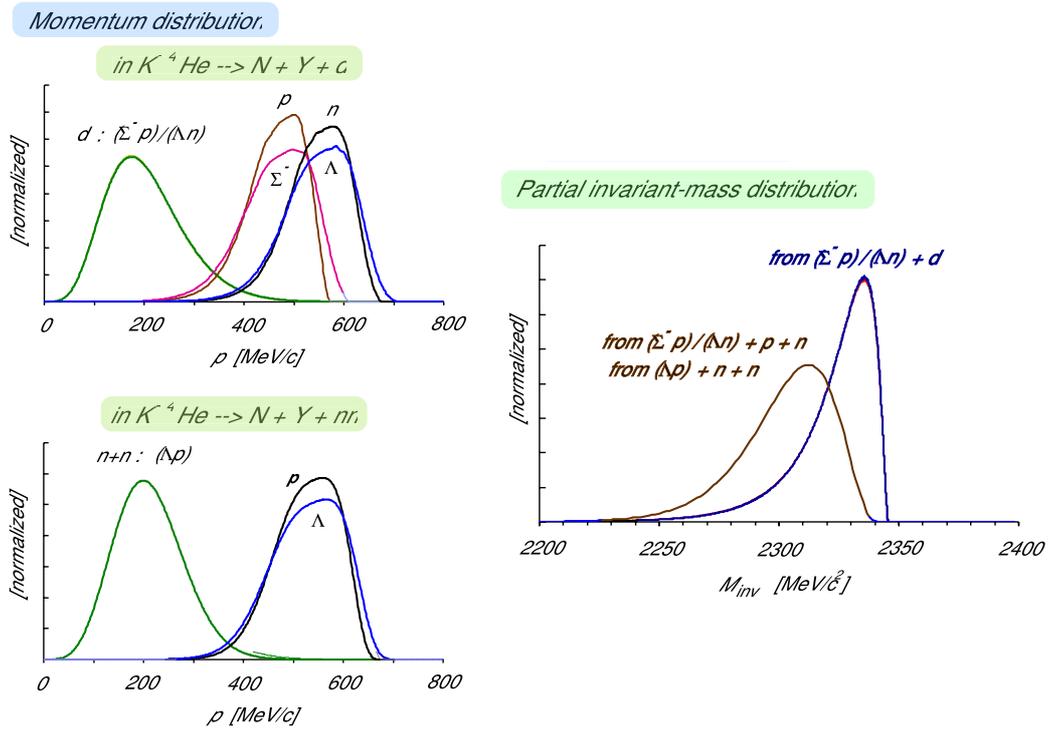}
\vspace{0.cm}
\caption{\label{fig:Summary} 
(Left) Summary of the calculated momentum distributions of emitted nucleons and hyperons for different residual systems in the $K^- + ^4$He absorption at rest. (Right) Partial invariant-mass spectra of $YN$. Only the Direct QF processes (B) are taken into account without FSI. }
\end{figure}

As a summary we show in Fig.~\ref{fig:Summary} the emission and partial invariant-mass spectra for all the combinations of $Y$-$N$ pairs with residual systems of $d$, $pn$ and $nn$. At this stage no effect of FSI is taken into account. The FSI is expected to take some fraction (less than 50 \%) of the $M_{\rm inv} (YN)$ peaks to the lower mass region, producing a broad background, but not a peak. 

Detailed experimental studies of the emission and invariant mass spectra are eagerly waited for to understand the absorption mechanism and to search for exotic kaonic bound states (C) beyond the process (B).  All the interesting experimental observations discussed here should require further confirmations by forthcoming experiments at KEK, FINUDA and AMADEUS.\\

The authors would like to thank Professor Paul Kienle for stimulating discussion. This work is supported by Grant-in-Aid by Monbu-Kagaku-Sho. One of the authors (T. Y.) is grateful to the support of the Alexander von Humboldt Foundation through its ``Forschungspreis".


\newpage

\end{document}